\documentclass[aps,pra,twocolumn,superscriptaddress,nofootinbib]{revtex4-2}
\usepackage{amsmath, amssymb}
\usepackage{mathrsfs}
\usepackage{array}
\usepackage{booktabs}
\usepackage{tabu}
\usepackage{dcolumn}
\usepackage{amsfonts}
\usepackage{float}
\usepackage{siunitx}
\usepackage{graphicx,color}
\usepackage[colorlinks={true}]{hyperref}
\hypersetup{colorlinks=true,linkcolor=red,citecolor=blue,urlcolor=blue}
\usepackage{bm}
\usepackage{subcaption}
\usepackage{pstricks}
\usepackage{braket}
\graphicspath{{images/}}
\usepackage{caption}
\expandafter\let\csname equation*\endcsname\relax
\expandafter\let\csname endequation*\endcsname\relax
\usepackage{mathtools}
\usepackage[utf8]{inputenc}
\usepackage[T1]{fontenc}
\usepackage{mathptmx}
\usepackage[markup=underlined]{changes}
\usepackage{cleveref}
\usepackage{etoolbox}
\usepackage{multirow}
\usepackage{orcidlink}
\usepackage{listings}
\usepackage{xcolor}

\begin{document}

\title{Photon Demultiplexing Using Spatial Routing}

\author{Hashir~Kuniyil\orcidlink{0000-0003-0338-1278}} 
\email{hkuniyil@hbku.edu.qa}
\affiliation{Qatar Center for Quantum Computing, College of Science and Engineering, Hamad Bin Khalifa University, Doha, Qatar}
\author{Asad Ali\orcidlink{0000-0001-9243-417X}} 
\affiliation{Qatar Center for Quantum Computing, College of Science and Engineering, Hamad Bin Khalifa University, Doha, Qatar}
\author{Saif Al-Kuwari\orcidlink{0000-0002-4402-7710}}
\affiliation{Qatar Center for Quantum Computing, College of Science and Engineering, Hamad Bin Khalifa University, Doha, Qatar}

\date{\today}

\begin{abstract}
Spatial sectioning of the far-field noncollinear field from spontaneous parametric down-conversion (SPDC) is a promising approach for multiparty quantum communication. Although demonstrations with six pairs of users have been reported, the fundamental limits remain unknown. These include the maximum number of channels, the required degree of channel optimization, and the dependence on input parameters. In this paper, we provide a practical framework that answers these questions by connecting input parameters and invoking a design theorem. For a representative configuration, we realize 12 user-pair channels with routing consistency exceeding $0.9999$ at a $12\%$ reduction in photon flux. We further show that sectioning improves the accidental-coincidence budget linearly with the number of channels $N$, allowing the aggregate key rate to scale with the user count. These results provide a design strategy for spatial-division multiplexing of entangled photons analogous to the channel-spacing standards used in wavelength-division networks.
\end{abstract}

\maketitle
\section{Introduction}

Distributing entanglement from a single source to many user pairs is the defining resource problem of quantum networks. In this, every added user pair only costs an additional receiver, not a source~\cite{Wengerowsky2018, Joshi2020, Sidhu2021}. Among the known multiplexing strategies, passive space-division multiplexing (SDM) of the emission of bulk spontaneous parametric down-conversion (SPDC) is uniquely economical and compact. An efficient segmentation strategy would allow a single source to serve multiple pairs of independent users on a QKD network. The core of this technique is the transverse-momentum conservation that places the two photons of each pair at diametrically opposite points of the emission ring, so a wedge introduced in the far field and extracting its antipodal partner form a self-contained two-user entanglement channel, selected by geometry alone, without spectral filters, modulators, and mode transformations on the photon path. Here, the receiver hardware is fundamentally determined by the spatial and functional placement of the detectors.

In a previous work, the SDM technique has been used to demonstrate a polarization encoded QKD system where spatial sectioning is performed through a specialized multicore fiber \cite{Ortega2021PRXQuantum}. This technique is an efficient way of extracting the independent entanglement modes conceptually equivalent to the passive division of the SPDC noncollinear ring into multiple independent bipartite state similar to the demonstration in which prism or D-shaped mirrors were used for sectioning~\cite{nai2025beam, nai2025free}. The same principle is used that served as parallel sources for device-independent randomness generation~\cite{nai2025device}, azimuthal sectioning underlies proposals for multiplexed heralded single-photon sources~\cite{Migdall2002, Pittman2002}. Yet in every one of these demonstrations, the segmentation itself is arbitrary in the sense that the number of segments, their widths, and the gaps between them are set by the available hardware including sectioning optical components and the fiber-core layout, rather than by source property of the biphoton state. The questions that define the technology are, to the best of our knowledge, open and important for advancement. In this work, we seek to answer defining questions: How many segments can a given source support at a given routing fidelity? What does each segment cost in the photon budget? Where should the boundaries be placed and how should the gaps between segments be sized? 
%Currently, popular field divion technique such as wavelength division multiplexing (WDM) became a network technology only when channel spacing, crosstalk, and guard bands were standardized against the physics of the medium~\cite{ITUgrid}. There is no corresponding standardization for SDM of entangled photons. Notice that many the method provided in this work is applicable only in SPDC generation in a bulk crystal configuration with distinguishable nonlinear ring is available to spatial sectioning.

The physics needed to answer these questions is, individually, well established. The transverse correlations of SPDC are comprehensively characterized~\cite{Walborn2010, Howell2004, Ling2008}, including full-field camera-based measurements of the momentum anti-correlation~\cite{Edgar2012, Moreau2014}. The azimuthal sector of the entanglement has been quantified as a resource. Fedorov derived the azimuthal Schmidt decomposition of noncollinear biphotons and proposed a multichannel Schmidt-type expansion~\cite{Fedorov2016azimuthal, Fedorov2004}, the conjugate orbital-angular-momentum description is captured by the spiral bandwidth~\cite{LawEberly2004, Torres2003spiral, Miatto2011}, and the angular Schmidt spectrum has been measured directly~\cite{Kulkarni2018angular}. Deriving network design rules from these established fundamentals is desirable, given the suitability of SDM for multi-user QKD \cite{Ortega2021PRXQuantum, nai2025beam, nai2025free}, the work is important for the efficient execution of the multi party QKD communication. We establish a closed-form relation between wedge count and routing fidelity, photon-budget accounting per segment, and optimal segmentation strategy.

The developed closed form formula is verified against a full pair-level Monte Carlo simulation of the joint biphoton amplitude. We find that the only decorrelation between a photon's azimuth and its partner's is the transverse pump spreading, so the azimuthal routing kernel is exactly $\sigma_\varphi = 1/(w_0 q_0)$, with $w_0$ the pump waist and $q_0$ the radius of noncollinear annulus in transverse momentum. Cylindrical symmetry then makes the boundary-crossing statistics exact, giving the consistency of per-channel routing for $N$ antipodal wedge-pair channels, $1-\eta(N) = \sqrt{2/\pi}\,\sigma_\varphi N/\pi$, where $\eta$ is the fidelity of channel routing, and hence the capacity law $N_{\max}(\eta)\approx3.94(1-\eta)\,w_0q_0$, which answers the central design question how many channels a given source supports at a given fidelity. Thus, every design question converges into the single dimensionless parameter $w_0q_0$, the radius of the ring measured in pump-waist units. Notably, the capacity is independent of the crystal length, which sets only the ring thickness and brightness. The control parameters that govern the space-division capacity are orthogonal to those that govern the source rate.

Three consequences give the developed framework its reach. First, expressing $w_0 q_0$ through the azimuthal Schmidt number $K_{\mathrm{az}}=\sqrt{\pi}\,w_0 q_0$ produces a resource theorem, $N_{\max}(\eta) \approx 2.2\,(1-\eta)\,K_{\mathrm{az}}$. This means that the number of usable spatial channels is a fidelity-budget fraction of the azimuthal entanglement resource, making the Schmidt analyzes of Refs.~\cite{Fedorov2016azimuthal, Kulkarni2018angular} directly operational. Second, we prove a guard-band theorem, i.e dead zones of width $g$ at the wedge boundaries suppress routing errors exponentially, as $\sim e^{-2g^2/\sigma_\varphi^2}$, at only linear photon-budget cost. The trade is favorable in practice, in a representative configuration, 12 user-pair channels reach routing consistency above $0.9999$ for a $12\%$ flux sacrifice, and it converts segmentation from an act of hardware convenience into a one-parameter optimization with a provably favorable scaling. Third, the photon budget is exact and automatically fair ($1/N$ per channel by symmetry), and sectioning improves the signal-to-accidental ratio per channel, because cross-segment accidentals fail the antipodal coincidence condition; the accidental-coincidence scaling improves with the number of channels. 

Although the azimuthal Schmidt modes are orbital-angular-momentum (OAM) waves, the present scheme is not OAM multiplexing; OAM encoding operates on the conjugate basis and requires mode sorters or projective holograms with loss per-mode~\cite{Berkhout2010Efficient,mirhosseini2013efficient, Torres2003spiral, Miatto2011}, while wedge sectioning is detector-native, the position basis that the camera or fiber array already measures. It is not high-dimensional "pixel entanglement" imaging~\cite{Edgar2012, Moreau2014}, which quantifies the dimensionality of the joint state rather than the fidelity and budget of routing \emph{pairs of users}. And it is complementary to wavelength-division approaches~\cite{Wengerowsky2018, Joshi2020}: at degenerate operation, the angular dispersion vanishes and the emission carries no spectral labels, so the azimuthal position is precisely the passive multiplexing resource that survives.

The paper is organized as follows. Section~\ref{sec:geometry} establishes the geometry and derives the azimuthal routing kernel from the joint biphoton amplitude. Section~\ref{sec:capacity} derives the exact per-channel routing consistency and the capacity law and verifies both by Monte Carlo simulation. Section~\ref{sec:guard} develops the guard-band optimization and the resulting fidelity--budget trade-off curves. Section~\ref{sec:budget} treats the per-segment photon budget, the polarization-fidelity uniformity across segments, and the accidentals scaling, assembling the per-channel secret-key budget for wavelength-independent multi-user BBM92~\cite{BBM92, Kwiat1999, Anwar2021}. Section~\ref{secVI} summarizes the practical implications of the results. Section~\ref{sec:conclusion} discusses the contribution of the paper and its extended applications and concludes the work.

\section{Source geometry and the azimuthal routing kernel}
\label{sec:geometry}

\subsection{Joint amplitude and ring geometry}
\label{sec:jointamp}

We consider degenerate SPDC ($\lambda_s=\lambda_i\equiv\lambda_d=2\lambda_p$) in a nonlinear crystal of length $L$, pumped by a monochromatic Gaussian beam of waist $w_0$. In the paraxial, undepleted-pump limit, the joint transverse amplitude is the standard product of pump envelope and phase-matching function~\cite{Walborn2010},
\begin{equation}
A(\mathbf{q}_s,\mathbf{q}_i)
 = E_p(\mathbf{q}_s+\mathbf{q}_i)\,
   \mathrm{sinc}\!\left(\frac{\Delta k_z\,L}{2}\right),
\qquad
E_p(\mathbf{q}) = \exp\!\left(-\frac{w_0^2 q^2}{4}\right),
\label{eq:JSA}
\end{equation}
with transverse momenta $\mathbf{q}_{s,i}$ and longitudinal mismatch, to
second paraxial order,
\begin{equation}
\Delta k_z = D_0
 + \frac{|\mathbf{q}_s+\mathbf{q}_i|^2}{2k_p}
 - \frac{q_s^2}{2k_d} - \frac{q_i^2}{2k_d},
\qquad k_d = \frac{2\pi n(\lambda_d)}{\lambda_d}.
\label{eq:mismatch}
\end{equation}
The collinear detuning $D_0$ (set by the poling period, phase-matching angle, or crystal temperature) selects the emission geometry: $D_0=0$ gives the collinear disk, while $D_0>0$ produces an annulus peaked at the transverse radius
\begin{equation}
q_0 = \sqrt{D_0\, k_d}, \qquad \theta_0 = q_0/k_d,
\label{eq:ring}
\end{equation}
of the noncollinear SPDC annulus of half-opening angle $\theta_0$ (see Fig.\ref{fig:kernel-geometry}). Figure. \ref{fig:kernel-geometry} shows a schematic of the standard settings to generate the noncollinear SPDC, with the generated spatial field structure of the noncollinear annulus shown. We note that the framework applies specifically to bulk-crystal SPDC, where the emission forms a free-space ring suitable for spatial sectioning; waveguide-based sources, which confine the down-converted light to guided modes, do not produce such a ring and are outside the scope of this work.

We define two conventions. First, the pump envelope in Eq.~\eqref{eq:JSA} is an \emph{amplitude}; the corresponding intensity distribution of the pair-sum coordinate $\boldsymbol{\delta}\equiv\mathbf{q}_s+\mathbf{q}_i$ is Gaussian with per-axis standard deviation
\begin{equation}
\sigma_\delta = \frac{1}{w_0}.
\label{eq:sigmadelta}
\end{equation}
Second, we write $\mathcal{Q}(x)=\tfrac12\,\mathrm{erfc}(x/\sqrt2)$ for the standard Gaussian tail probability.

\subsection{Pair correlation in polar coordinates}
\label{sec:polar}

\begin{figure*}[htbp]
    \centering
    \includegraphics[width=1\linewidth]{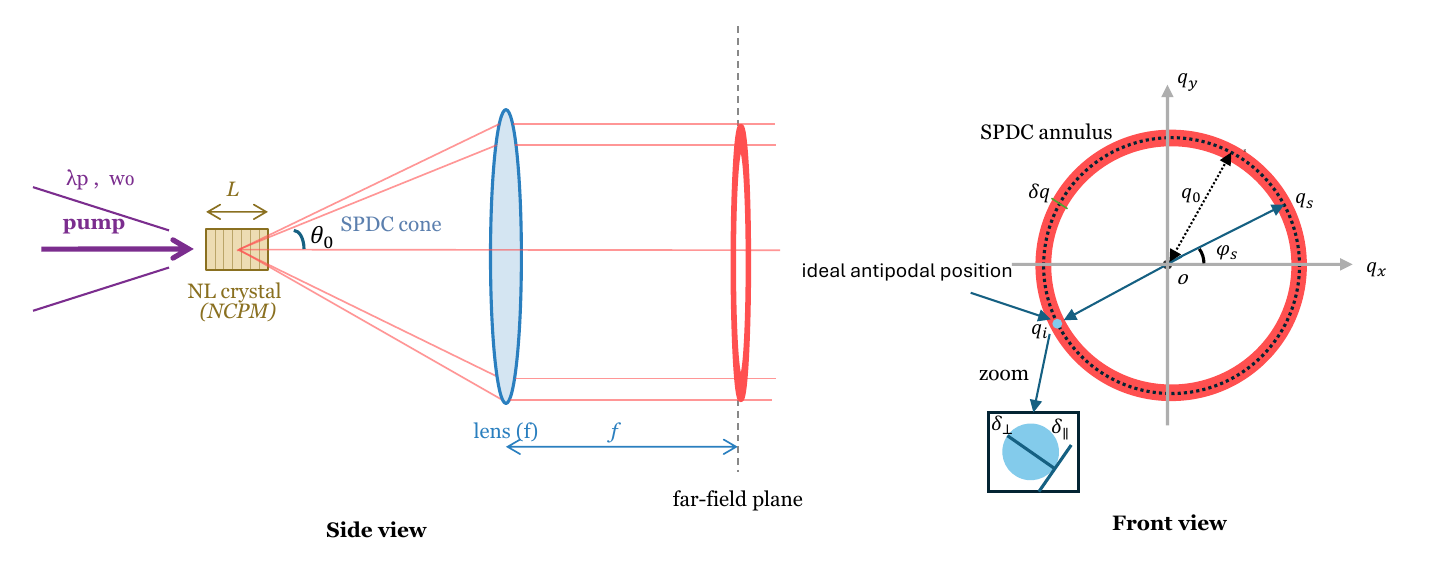}
    \caption{Schematic of the azimuthal routing kernel of a noncollinear SPDC source. \textbf{Side view:} a Gaussian pump beam (waist $w_0$, wavelength $\lambda_p$) impinges on a noncritically phase-matched (NCPM) in a noncollinear SPDC configuration. A collection lens of focal length $f$ maps the emission cone onto the far-field plane, which is observed as annular field. \textbf{Front view:} the far-field SPDC annulus of momentum central radius $q_0$ and radial thickness $\delta q$; signal and idler photons of a pair occupy antipodal points on the ring. The signal azimuth is $\varphi_s$; momentum conservation alone would place the idler at the exact antipode ($\times$~marker), but the finite pump waist displaces it by $\boldsymbol{\delta}\sim\mathcal{N}(0,\sigma_\delta^2\mathbb{I}_2)$ with $\sigma_\delta=1/w_0$ [Eq.~\eqref{eq:anticorr}]. \textbf{Zoom:} decomposition of $\boldsymbol{\delta}$ into components parallel ($\delta_\parallel$, radial) and perpendicular ($\delta_\perp$, azimuthal) to the signal direction. The radial component shifts only the common ring radius [Eq.~\eqref{eq:radial}]; the azimuthal component alone deflects the idler off the antipode by $\Delta\varphi=\delta_\perp/r_s$ [Eq.~\eqref{eq:azimuthal}], producing the Gaussian routing kernel of width $\sigma_\varphi=\sigma_\delta/q_0=1/(w_0q_0)$ [Eq.~(8)]. The ratio $\sigma_\delta/q_0$ is exaggerated for clarity; in every configuration considered, $\sigma_\varphi\lesssim10^{-2}$~rad. }
    \label{fig:kernel-geometry}
\end{figure*}

Transverse-momentum conservation, enforced by the pump envelope, constrains the pair as
\begin{equation}
\mathbf{q}_i = -\mathbf{q}_s + \boldsymbol{\delta},
\qquad \boldsymbol{\delta}\sim
\mathcal{N}\!\left(0,\sigma_\delta^2\,\mathbb{I}_2\right),
\label{eq:anticorr}
\end{equation}
so that, in ideal conditions where $\boldsymbol{\delta}=0$, the idler appears at the exact antipode of the signal on the ring (see the SPDC noncollinear ring shown in Fig.\ref{fig:kernel-geometry}). Write the signal in polar coordinates, $\mathbf{q}_s=r_s(\cos\varphi_s,\sin\varphi_s)$, and decompose $\boldsymbol{\delta}$ into components parallel and perpendicular to the
signal radial direction, $\boldsymbol{\delta} = \delta_\parallel\,\hat{\mathbf{q}}_s + \delta_\perp\,\hat{\mathbf{q}}_s^{\perp}$. To first order in $\delta/q_0$,
\begin{align}
r_i &= r_s - \delta_\parallel ,
\label{eq:radial}\\[2pt]
\varphi_i &= \varphi_s + \pi + \Delta\varphi,
\qquad
\Delta\varphi = \frac{\delta_\perp}{r_s} .
\label{eq:azimuthal}
\end{align}
Equations \eqref{eq:radial}--\eqref{eq:azimuthal} separate the pair correlation into its two polar channels, and the separation is the physical foundation of everything that follows:

\emph{(i) The radial channel is the common mode.} The two photons share their ring radius up to the spreading $\delta_\parallel$; both the intrinsic ring thickness and the radial spread displace the pair \emph{together} in radius. A receiver whose segments span the full radial extent of the annulus (for example, a ``bucket'' wedge integrating radially) is therefore insensitive to the radial channel: it contributes no routing information and no routing error. This choice also removes any mode mismatch—every radial mode within the wedge is collected—so the segmentation causes no heralding degradation relative to an unsegmented receiver.

\emph{(ii) The azimuthal channel carries a single Gaussian relative coordinate.} The relative azimuth of the pair, referred to the antipodal frame, is
Gaussian,
\begin{equation}
\Delta\varphi \sim \mathcal{N}\!\left(0,\sigma_\varphi^2\right),
\qquad
\sigma_\varphi = \frac{\sigma_\delta}{q_0} = \frac{1}{w_0\,q_0},
\label{eq:kernel}
\end{equation}
where $\sigma_\delta=1/w_0$ is the per-axis width of Eq.~\eqref{eq:sigmadelta} and we set $r_s\simeq q_0$ (thin-ring approximation). This is the \emph{azimuthal routing kernel}: the conditional distribution of the idler azimuth given the signal's, referred to the antipodal frame. Its width $\sigma_\varphi$ is the azimuthal analog of the conditional width (Fedorov) of the transverse correlations~\cite{Fedorov2004, Walborn2010}. Corrections to Eq.~\eqref{eq:kernel} arise from the finite ring thickness and the small-rotation expansion; because first-order radial variations of $\sigma_\delta/r$ cancel under the
ring average (contributions from $r>q_0$ and $r<q_0$ enter with opposite
sign), the leading correction is $O\!\left[(\delta q/q_0)^2\right]$, together with $O(\sigma_\delta/q_0)$ from the linearization of Eq.~\eqref{eq:azimuthal}. Both are small in the regime of every experiment considered here, and the residual effect is directly visible in the Monte Carlo of Sec.~\ref{sec:mc}: at $\theta_0=1^\circ$, where $\delta q/q_0\approx0.5$, the closed form and simulation differ by only $\sim10^{-3}$.

\subsection{Properties of the kernel: the single design parameter}
\label{sec:kernelprops}

Three properties of Eq.~\eqref{eq:kernel} organize the entire design space.
\emph{Crystal-length independence.} The length $L$ affects the joint amplitude only through the sinc function, i.e.\ only radially: it sets the annular thickness and the source brightness, but it does not affect $\sigma_\varphi$. The parameters that govern spatial-division capacity
($w_0$, $q_0$) are orthogonal to those that govern the pair rate ($L$, pump power), so brightness and capacity can be optimized independently.

\emph{Wavelength independence.} Degenerate noncollinear emission is broadband; each spectral pair $(\lambda_d+\Delta, \lambda_d-\Delta)$ occupies its own ring radius $r(\Delta)$, which is the origin of the spatio-spectral X structure analyzed \cite{Gatti2009PRL}. But every
spectral ring is azimuthally uniform and antipodally paired, and the kernel $\sigma_\varphi = \sigma_\delta/r$ varies only at second order across the occupied radii. Radially integrating wedges are therefore spectrally insensitive: the routing consistency derived below holds for the full SPDC bandwidth without spectral filtering. This contrasts with the nondegenerate regime, where the strong spatio-spectral correlation makes the emission angle a frequency label \cite{Kuniyil2026Entanglement}. In this sense, degenerate operation is the natural regime for SDM \cite{kuniyil2026spectral}.

\emph{One dimensionless parameter;} All segmentation questions reduce to
\begin{equation}
w_0 q_0 \;=\; \frac{\text{ring radius in momentum}}{\text{pump momentum
width}} ,
\label{eq:w0q0}
\end{equation}
of the radius of the ring measured in pump-waist units. Configurations with equal $w_0 q_0$ have identical azimuthal statistics regardless of how the product is realized, a scaling invariance that we verify numerically in Sec.~\ref{sec:mc}. Representative values: at $\lambda_d = 810$~nm ($k_d = 13.6\,\mu\mathrm{m}^{-1}$), a ring of $\theta_0=3^\circ$ pumped at $w_0 = 200\,\mu$m gives $w_0 q_0 \approx 143$ ($\sigma_\varphi \approx 0.40^\circ$ of azimuth); widening the pump to $600\,\mu$m gives $w_0 q_0 \approx 428$. These parameters are representative of noncritically phase-matched (NCPM)
KTP at $\lambda_d=810$~nm; the framework applies equally to
$\beta$-BBO type-I sources at visible wavelengths, where typical ring angles $\theta_0\sim3^\circ$--$5^\circ$ at pump waists $w_0\sim100$--$300\,\mu$m
give $w_0q_0$ in the same range.

\section{routing consistency and the channel-capacity law}\label{sec:capacity}

\begin{figure*}[htbp]
    \centering
    \includegraphics[width=0.8\textwidth]{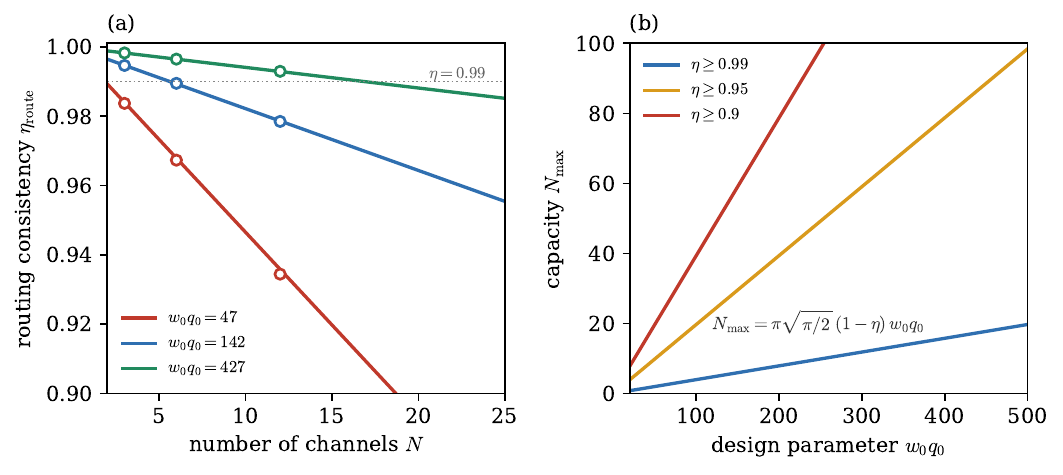}
    \caption{\label{fig:capacity}Per-channel routing consistency and the channel-capacity law. (a)~Routing consistency $\eta_{\rm route}$ versus the number of antipodal wedge-pair channels $N$. Solid curves are the closed form [Eq.~\eqref{eq:etaN}] for three representative rings,$(\theta_0,w_0)=(1^\circ,200\,\mu\mathrm{m})$, $(3^\circ,200\,\mu\mathrm{m})$, $(3^\circ,600\,\mu\mathrm{m})$, i.e.\ $w_0q_0=47,\,142,\,427$; open circles are pair-level Monte Carlo sampling the full joint biphoton amplitude ($\lambda_d=810\,$nm, $L=3\,$mm; $1.5\times10^{6}$ pairs per point) with no azimuthal idealization. Closed form and simulation agree to $\sim\!10^{-3}$ at every point; the dotted line marks $\eta=0.99$. (b)~Channel capacity $N_{\max}$ as a function of the single design parameter $w_0q_0$ at routing targets $\eta\geq0.99,\,0.95,\,0.90$ [Eq.~\eqref{eq:capacity}]. The capacity is linear in $w_0q_0$ and in the fidelity budget $(1-\eta)$, and is independent of crystal length.}
\end{figure*}
\subsection{Segmentation and per-channel routing consistency} \label{sec:routing}

Divide the azimuth of the far-field into $M = 2N$ contiguous wedges of equal width $\Delta\Phi = 2\pi/M = \pi/N$, and define the channel $c\in\{1,\dots,N\}$ as the pair of antipodal wedges $(W_c, W_{c+N})$: user $A_c$ collects wedge $W_c$, user $B_c$ the antipodal wedge. A pair is \emph{routed consistently} when the signal falls in some $W_c$ and the idler falls in its antipodal partner. By Eq.~\eqref{eq:azimuthal}, this reduces to a one-dimensional question: mapping the idler to the antipodal frame, consistency fails if and only if the relative coordinate $\Delta\varphi$ carries the idler across a boundary of the signal's wedge.

By cylindrical symmetry, the signal azimuth is uniformly distributed within its wedge; there is no ridge structure, no flux gradient, so the average requires no weighting. The boundary-crossing probability is therefore an unweighted average over the
wedge of the two Gaussian tail integrals, and it is \emph{exact}:
\begin{align}
1-\eta(N)
&= \frac{1}{\Delta\Phi}\int_0^{\Delta\Phi}
   \Big[\mathcal{Q}\!\big(\tfrac{u}{\sigma_\varphi}\big)
   + \mathcal{Q}\!\big(\tfrac{\Delta\Phi-u}{\sigma_\varphi}\big)\Big]du
\nonumber\\[2pt]
&= \frac{2\sigma_\varphi}{\Delta\Phi}
   \left[\frac{1}{\sqrt{2\pi}}\Big(1-e^{-X^2/2}\Big) + X\,\mathcal{Q}(X)\right],
\label{eq:etaexact}
\end{align}
where $X \equiv \frac{\Delta\Phi}{\sigma_\varphi}$, the second line uses $\int_0^{X}\mathcal{Q}(v)\,dv = X\mathcal{Q}(X) + \frac{1}{\sqrt{2\pi}}(1-e^{-X^2/2})$. In the operating regime $\Delta\Phi\gg\sigma_\varphi$ the correction terms in Eq.~\eqref{eq:etaexact} are exponentially small [$X\gtrsim30$ in the configurations of Table~\ref{tab:mc}], the bracket approaches $1/\sqrt{2\pi}$, and Eq.~\eqref{eq:etaexact} collapses to the working form 
\begin{equation}
1-\eta(N) \;=\;\frac{2\sigma_\varphi}{\Delta\Phi}\frac{1}{\sqrt{2\pi}}
\;=\; \sqrt{\frac{2}{\pi}}\;\frac{\sigma_\varphi\,N}{\pi}
\label{eq:etaN}
\end{equation}
per channel, identical for every channel. Equivalently, $1-\eta=\mathbb{E}|\Delta\varphi|/\Delta\Phi$ with $\mathbb{E}|\Delta\varphi|=\sigma_\varphi\sqrt{2/\pi}$ the mean absolute deviation of the kernel: the routing loss is the mean azimuthal wander of a pair measured in wedge widths. Two remarks. First, Eq.~\eqref{eq:etaN} counts disagreements, events in which exactly the relative coordinate, not the pair, crosses a boundary. A pair displaced as a whole (by ring thickness, by the common-mode radial smearing, or simply by where on the ring it was born) lands intact in some channel and is a valid event for that channel; only the relative coordinate produces an unmatched single, removed at sifting. Routing failure is therefore a rate loss, not an error mechanism, which accounts for the separation of this framework from a crosstalk-QBER model. Second, leakage into non-adjacent wedges is suppressed as $e^{-(\Delta\Phi/\sigma_\varphi)^2/2}$ and is negligible at every parameter set considered here. Figure~\ref{fig:capacity}(a) shows the routing consistency as a function of the number of channels for three values of $w_0q_0$: a larger $w_0q_0$ supports more channels at lower routing loss. Open circles are pair-level Monte Carlo results at $N=3$, $6$, and $12$, which agree with the closed form to $\sim10^{-3}$.
\subsection{The capacity law}
\label{sec:law}

Rearranging Eq.~\eqref{eq:etaN} at a routing target $\eta$ gives the channel capacity of the source,
\begin{equation}
N_{\max}(\eta) \;=\;
\left\lfloor \pi\sqrt{\frac{\pi}{2}}\;(1-\eta)\;w_0 q_0 \right\rfloor
\;\approx\; \big\lfloor 3.94\,(1-\eta)\, w_0 q_0 \big\rfloor.
\label{eq:capacity}
\end{equation}
The capacity is linear in the fidelity budget $(1-\eta)$ and linear in the single design parameter $w_0 q_0$; it inherits the crystal-length
independence and wavelength independence of the kernel. Numerically, the $\theta_0 = 3^\circ$, $w_0=200\,\mu$m ring supports $N_{\max}=5$ channels at
$\eta \ge 0.99$ and $28$ at $\eta\ge 0.95$; widening the pump to
$600\,\mu$m raises these to $16$ and $84$. Figure~\ref{fig:capacity}(b) shows the channel capacity as a function of $w_0q_0$ at three
routing targets $\eta$. We emphasize that
Eq.~\eqref{eq:capacity} replaces the segment count, which was a hardware decision (as in every published sectioning experiment); here it is a source property, computable from two measured numbers before any sectioning is employed.

\subsection{Resource theorem: capacity as a fraction of the azimuthal
Schmidt number}
\label{sec:resource}
The capacity law acquires its information-theoretic meaning when expressed
against the azimuthal entanglement: the number of usable wedge-pair
channels at routing fidelity $\eta$ is a fidelity-budget fraction of the
azimuthal Schmidt number $K_{\rm az}$,
\begin{equation}
N_{\max}(\eta)\approx\frac{\pi}{\sqrt{2}}\,(1-\eta)\,K_{\rm az}
\approx 2.2\,(1-\eta)\,K_{\rm az}.
\label{eq:resource}
\end{equation}
We derive $K_{\rm az}$ and show how this follows from the capacity law. Tracing the radial coordinates of Eq.~\eqref{eq:JSA} in the thin-ring limit leaves a
translationally invariant azimuthal amplitude kernel $g(\varphi_s-\varphi_i-\pi)$, Gaussian with amplitude-level width $\sigma_A = \sqrt{2}\,\sigma_\varphi$ (the intensity kernel of Eq.~\eqref{eq:kernel} being its square). Translational invariance on the circle makes the Schmidt modes exactly the orbital-angular-momentum (OAM) waves $e^{im\varphi}$~\cite{mair2001entanglement}, with Schmidt coefficients given by the Fourier coefficients of $g$: $\lambda_m \propto e^{-m^2\sigma_A^2}$, a Gaussian spiral spectrum~\cite{Torres2003spiral, Miatto2011}. The azimuthal Schmidt
number follows in closed form,
\begin{equation}
K_{\mathrm{az}} = \frac{\big(\sum_m \lambda_m\big)^2}{\sum_m \lambda_m^2}
\;\simeq\; \frac{\sqrt{2\pi}}{\sigma_A}
\;=\; \frac{\sqrt{\pi}}{\sigma_\varphi}
\;=\; \sqrt{\pi}\; w_0 q_0 ,
\label{eq:Kaz}
\end{equation}
recovering, in our geometry, the pump-waist scaling of Fedorov's azimuthal Schmidt analysis~\cite{Fedorov2016azimuthal} and the measured angular Schmidt spectra of Ref.~\cite{Kulkarni2018angular}. The finite width of the spiral spectrum bounds the populated modes to $|m|\lesssim1/\sigma_A$; $K_{\rm az}$ is the participation number of this distribution. Substituting Eq.~\eqref{eq:Kaz} into Eq.~\eqref{eq:capacity}, recovers Eq.~(\eqref{eq:resource}). We stress the division of work with the prior literature: Eq.~\eqref{eq:Kaz} quantifies a resource and is, in substance, known~\cite{Fedorov2016azimuthal, Kulkarni2018angular, LawEberly2004}; Eq.~\eqref{eq:resource} is the \emph{operational} result,it converts that resource into a segment count at a stated fidelity, which is experimentally verifiable. It also exposes the tradeoff of the position-basis simplicity: wedge sectioning accesses $\sim 2(1-\eta)$ of the Schmidt modes as independent channels, whereas a full OAM-sorting receiver could, in principle, address all $K_{\mathrm{az}}$ at the cost of mode-transformation hardware and per-mode loss \cite{Berkhout2010Efficient, mirhosseini2013efficient}; the wedge architecture trades Hilbert-space efficiency for detector-native passivity. (Equation~\eqref{eq:Kaz} idealizes the ring as thin and the pump as Gaussian; finite ring thickness and spectral breadth multiply the \emph{total} spatial Schmidt number by radial factors but leave the azimuthal factor, and hence Eq.~\eqref{eq:resource}, intact.)
\subsection{Monte Carlo verification}
\label{sec:mc}

We verify Eqs.~\eqref{eq:etaN}--\eqref{eq:capacity} against pair-level Monte Carlo sampling of the full joint amplitude, with no azimuthal idealizations: signal momenta are sampled from the exact annular distribution $p(r)\propto r\,\mathrm{sinc}^2[(D_0 - r^2/k_d)L/2]$ at $L=3$~mm with uniform azimuth, the pump spread $\boldsymbol{\delta}$ from the two-dimensional Gaussian of Eq.~\eqref{eq:anticorr}, and the idler constructed vectorially as $\mathbf{q}_i=-\mathbf{q}_s+\boldsymbol{\delta}$ before wedge assignment, so the small-rotation expansion of
Eq.~\eqref{eq:azimuthal}, the thin-ring limit, and the uniform-density step of Eq.~\eqref{eq:etaexact} are all tested. Table~\ref{tab:mc} summarizes the comparison across ring angles $\theta_0\in\{1^\circ,2^\circ,3^\circ\}$, pump waists $w_0\in\{200,600\}\,\mu$m, and segment count $N=12$: closed form and simulation agree to within $2\times10^{-3}$ at every point; the small systematic offsets in the $\theta_0=1^\circ$ rows are the second-order ring-thickness corrections discussed in Sec.~\ref{sec:polar}. The scaling invariance of Eq.~\eqref{eq:w0q0} is verified directly: the configurations $(\theta_0,w_0)=(1^\circ,600\,\mu\text{m})$ and $(3^\circ,200\,\mu\text{m})$, which share $w_0q_0\approx 143$, return identical routing statistics to within sampling error despite a nine-fold difference in solid angle. The same simulation, with dead zones inserted at the wedge boundaries, generates the guard-band trade-off curves of Sec.~\ref{sec:guard}.

% ,,,,,,,,,,,,,,,,,,,,,,
\begin{table}
\caption{\label{tab:mc}Per-channel routing consistency $\eta$ at $N=12$: closed form [Eq.~\eqref{eq:etaN}] versus pair-level Monte Carlo ($10^{7}$ samples per point; statistical uncertainty $\lesssim1\times10^{-4}$). $\lambda_d=810$~nm, $L=3$~mm. The two starred rows share $w_0q_0\approx143$ and verify the single-parameter scaling of Eq.~\eqref{eq:w0q0}. The small deviations in the $\theta_0=1^\circ$ rows, where the relative ring thickness is largest ($\delta q/q_0\approx0.5$), are the second-order thickness corrections discussed in Sec.~\ref{sec:polar}.}

\begin{ruledtabular}
\begin{tabular}{ccccccc}
$\theta_0$ & $w_0$ ($\mu$m) & $w_0q_0$ & $K_{\rm az}$ & $N$ &
$\eta$ (closed form) & $\eta$ (MC) \\
\colrule
$1^\circ$ & 200      & 48  & 84  & 12 & 0.9358 & 0.9346 \\
$2^\circ$ & 200      & 95  & 169 & 12 & 0.9679 & 0.9677 \\
$3^\circ$ & 200$^*$  & 143 & 253 & 12 & 0.9786 & 0.9787 \\
$1^\circ$ & 600$^*$  & 143 & 253 & 12 & 0.9786 & 0.9781 \\
$3^\circ$ & 600      & 428 & 758 & 12 & 0.9929 & 0.9929 \\
\end{tabular}
\end{ruledtabular}
\end{table}
% ==================================================================
% Sections IV & V — azimuthal-sectioning framework (PRX Quantum)
% Follows sections_II_III_azimuthal.tex; keys match references_azimuthal.bib.
% NOTE: add DevetakWinter2005 and ShorPreskill2000 to the .bib (copy from
% the frequency-paper bibliography).
% Verified numbers (azimuthal_slicing.py, N=12, theta0=3deg, w0=200um,
% sigma_phi=0.40deg, DeltaPhi=15deg):
%   bare:        eta = 0.978-0.979, acceptance 1.00
%   guard  5%:   eta > 0.9999 (MC: no disagreements in ~7x10^5 accepted),
%                acceptance 0.88
%   guard 10%:   eta > 0.9999, acceptance 0.78
% Closed-form residual with guard: prefactor-exact expression below;
% leading suppression exp(-2 g^2/sigma_phi^2) (effective gap 2g since both
% photons must clear their guard zones).
% ==================================================================

\section{Guard-band optimization}
\label{sec:guard}

The capacity law of Eq.~\eqref{eq:capacity} does not account for the dead zones where photons are not collected, unregistered camera pixels, fiber cladding, seams between mirror segments, the gaps between multicore-fiber cores \cite{Ortega2021PRXQuantum}. This section shows that such dead zones, whether imposed by hardware or placed deliberately, are not a defect but the optimal use of the photon budget: they gain fidelity exponentially at a cost that is only linear, and we prove that boundary-centered guard zones are the optimal shape for the sacrificed region.

\subsection{Residual error with guard bands}
\label{sec:guardresidual}
\begin{figure}[htbp]
    \centering \includegraphics[width=\columnwidth]{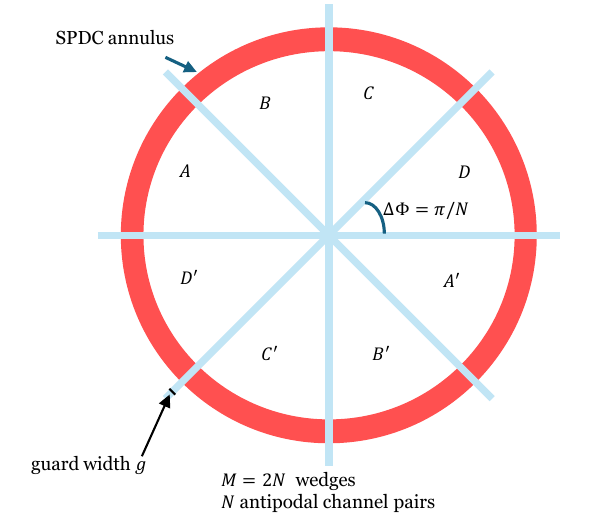}
    \caption{Wedge segmentation of the SPDC annulus into $N=4$ antipodal channel pairs. The far-field ring is divided into $M=2N=8$ equal wedges of angular width $\Delta\Phi=\pi/N$; each antipodal pair (e.g.\ $A/A'$, $B/B'$) forms one self-contained two-user entanglement channel. Light-shaded strips of width $g$ at every wedge boundary are the guard bands of Sec.~\ref{sec:guardresidual}: photons landing in these dead zones are discarded at the coincidence level, suppressing routing errors exponentially [Eq.~\eqref{eq:guardtheorem}] at a linear budget cost of $\approx 2g/\Delta\Phi$. The prescription uniform wedges $+$ guards sized by Eq.~\eqref{eq:budgetcost}} is the optimal segmentation of a degenerate SPDC ring (Sec.~\ref{sec:optimality}).
    \label{fig:guardBand}
\end{figure}

Assume a dead zone of width $g$ at each edge of each wedge as shown in Fig.~\ref{fig:guardBand}, so that each wedge retains a core of width $\Delta\Phi-2g$ and each boundary is bordered by a total gap of $2g$. An event is accepted only if both photons land in their respective cores; this is the natural coincidence-level acceptance, and it is what doubles the effective guard. A routing disagreement now requires the relative coordinate to carry the idler from inside the signal's core, across the boundary, and past the neighbor's guard zone: for a signal at distance $u\ge g$ from its nearest boundary, the crossing probability is $\mathcal{Q}\!\big((u+g)/\sigma_\varphi\big)$. Averaging over the uniform signal position in the core and summing the two boundaries gives exactly,
\begin{equation}
1-\eta(N,g)
= \frac{2\sigma_\varphi}{\Delta\Phi-2g}
  \left[\frac{e^{-2g^2/\sigma_\varphi^2}}{\sqrt{2\pi}}
        - \frac{2g}{\sigma_\varphi}\,
          \mathcal{Q}\!\left(\frac{2g}{\sigma_\varphi}\right)\right],
\label{eq:guardexact}
\end{equation}
obtained from $\int_{X}^{\infty}\mathcal{Q}(v)\,dv = \tfrac{1}{\sqrt{2\pi}}e^{-X^2/2} - X\mathcal{Q}(X)$ with $X=2g/\sigma_\varphi$. The effect of the dead zones is captured by the error-suppression ratio:
\begin{align}
\frac{1-\eta(N,g)}{1-\eta(N,0)} \;\simeq\; e^{-2g^2/\sigma_\varphi^2}
\quad\text{(fidelity gain: exponential)} \label{eq:guardtheorem},\\ 
1-\mathcal{A}(g) \;\simeq\; \frac{2g}{\Delta\Phi}
\quad\text{(budget cost: linear)}
\label{eq:budgetcost}
\end{align}
where $\mathcal{A}(g)$ is the acceptance of the coincidence (photon budget retained). Inverting the exponential, the guard required to suppress the residual error by a factor $S$ grows only as the square root of its logarithm,
\begin{equation}
g^\ast(S) = \frac{\sigma_\varphi}{\sqrt{2}}\,\sqrt{\ln S},
\qquad
\text{budget cost} \;=\; \frac{2g^\ast}{\Delta\Phi}
= \sqrt{2\ln S}\;\frac{\sigma_\varphi}{\Delta\Phi}.
\label{eq:guardsizing}
\end{equation}
A hundred-fold error suppression costs a guard of $1.5\,\sigma_\varphi$ per
edge; a ten-thousand-fold suppression costs $2.1\,\sigma_\varphi$. Because
$\sigma_\varphi/\Delta\Phi = \sqrt{\pi/2}\,(1-\eta_{\mathrm{bare}})$ is
already small in any sensible design, the budget prices are single-digit
percentages.

\begin{figure*}[htbp]
    \centering
    \includegraphics[width=0.8\textwidth]{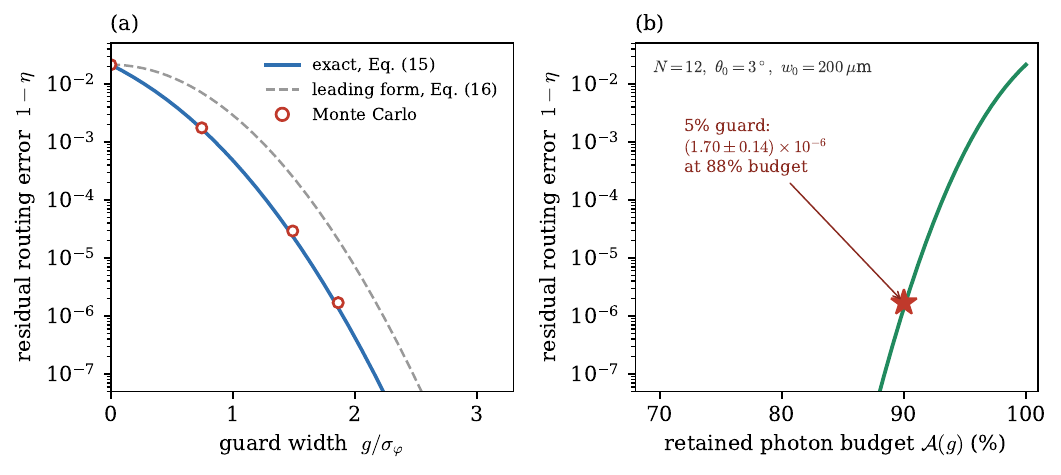}
    \caption{(a) Residual routing error $1-\eta$ versus the guard width $g$ inserted at each wedge boundary. The solid curve is the exact result, Eq.~\eqref{eq:guardexact}; the dashed curve is the leading form of Eq.~\eqref{eq:guardtheorem}, which overestimates the residual at large $g$ because the two terms of the bracket in Eq.~\eqref{eq:guardexact} nearly cancel. Circles are pair-level Monte Carlo with statistical error bars (logarithmic ordinate). (b) Residual error versus the retained photon budget $\mathcal{A}(g)$. The star marks a $5\%$ guard, at which the Monte Carlo residual is $(1.70\pm0.14)\times10^{-6}$ ($10^{8}$ pairs) while $88\%$ of the coincidences are retained.}
    \label{fig:guardband}
\end{figure*}

Monte Carlo, with the dead zones inserted directly into the wedge assignment of Sec.~\ref{sec:mc}, confirms the trade at its most demanding tested point ($N=12$, $\theta_0=3^\circ$, $w_0=200~\mu$m; bare $\eta=0.979$, $\sigma_\varphi=0.40^\circ$, $\Delta\Phi=15^\circ$): at a guard of $5\%$ of the wedge width ($g=1.9\,\sigma_\varphi$), $10^{8}$ simulated pairs yield a residual disagreement fraction of $(1.70\pm0.14)\times10^{-6}$, in agreement with the exact prediction $1.4\times10^{-6}$ of Eq.~\eqref{eq:guardexact}, while $88\%$ of the photon budget is retained; a $10\%$ guard retains $78\%$. Twelve user-pair channels at routing consistency exceeding $0.9999$, from a single $200$-$\mu$m-pumped ring, for a $12\%$ flux sacrifice. Figure~\ref{fig:guardband}(a) compares the exact residual of Eq.~\ref{eq:guardexact} with its leading form, Eq.~(16), and with Monte Carlo; Fig.~\ref{fig:guardband}(b) shows the resulting fidelity--budget trade-off, with the $5\%$-guard operating point marked.
\subsection{Optimality of boundary-centered guards}
\label{sec:optimality}

The guard-band strategy is not only effective; it is, in fact, optimal among all methods of allocating a fixed photon budget. Fix the acceptance measure (the total angular length of the accepted set within a wedge) and evaluate which accepted set minimizes the disagreement probability. The probability that an accepted signal at position $u$ (distance to the nearest boundary) produces a disagreement is monotonically decreasing in $u$: points near a boundary contribute strictly larger disagreement probability than points near the wedge center. Minimizing the average of a monotone function over a set of fixed measure means discarding the points where the function is largest,i.e., discarding contiguous neighborhoods of the boundaries. Boundary-centered guard zones are therefore the exact minimizer; no interleaved, apodized, or off-center sacrifice pattern can do better at equal budget. Combined with the symmetry argument that equal-width wedges are optimal for equal-rate channels (the azimuthal flux being uniform, any unequal partition only unbalances rates without improving any channel's kernel), the prescription \emph{uniform wedges + guards sized by Eq.~\eqref{eq:guardsizing}} is the optimal segmentation of a degenerate SPDC ring, closing the strategy question that the ad-hoc layouts of Refs.~\cite{Ortega2021PRXQuantum, nai2025beam, nai2025free} left open. The guards require no additional hardware: unclaimed camera pixels, fiber-cladding gaps, or mirror seams; in fact, existing experiments contain \emph{accidental} guard bands (fiber fill factors, D-mirror edges) whose sizes the framework can assess against Eq.~\eqref{eq:guardsizing}.

\section{Photon budget, polarization uniformity, and the per-channel key
budget}\label{sec:budget}

\subsection{Photon budget and rates}
\label{sec:photonbudget}

Cylindrical symmetry makes the budget accounting exact and automatically fair. Each of the $2N$ wedges collects the fraction $1/(2N)$ of all detected single photons, and each channel (wedge pair) collects the fraction $1/N$ of all coincidences; with guards, this becomes $\mathcal{A}(g)/N$. No photon count rate equalization step is needed,in contrast to segmentations of structured distributions, where boundary optimization must trade rate balance against fidelity, the annular ring shape gives equal rates by construction.

Because the wedges integrate over the full radial thickness and spectral breadth of the annulus (Sec.~\ref{sec:polar}), the collection carries no mode-selection penalty: the heralding chain is the unsegmented one, $\eta_h=\eta_{\rm opt}^2\eta_{\rm det}^2\eta_{\rm
coinc}$, and the segmentation is a pure relabeling of where each pair lands. The per-channel sifted-coincidence rate can be estimated using
\begin{equation}
R_c \;=\; \frac{\mathcal{A}(g)}{N}\;\eta(N,g)\;
R_{\mathrm{coinc}},
\label{eq:ratechain}
\end{equation}
with $R_{\mathrm{coinc}} = R_{\mathrm{pair}}\,\eta_h$ the total collected coincidence rate. The only geometry-specific optical requirement is
aperture: collecting a $\theta_0=3^\circ$ annular ring needs numerical aperture $\mathrm{NA}\gtrsim\sin\theta_0\approx 0.05$, trivially available, so the ring angle, which buys capacity linearly through $q_0$, is nearly free optically until aberrations of the collection lens intervene.

\subsection{Polarization uniformity across segments}
\label{sec:poluniformity}

For polarization-entangled ring sources, the delivered two-photon polarization state can vary with azimuth: noncollinear projection effects rotate the local polarization frames (entering at $O(\theta_0^2)$ and negligible at the few-degree rings considered here), while critically phase-matched bulk crystals and interferometric source geometries can imprint an azimuth-dependent relative phase $\chi(\varphi)$ on the $|HH\rangle$/$|VV\rangle$ superposition through walk-off and path-length asymmetries; quasi-phase-matched collinear-detuned sources and Sagnac geometries are azimuthally uniform by construction~\cite{Kim2006Phase, fedrizzi2007wavelength, Anwar2021}. The framework treats any smooth $\chi(\varphi)$ with a visibility argument: a constant phase per channel is harmless, because each user pair absorbs the wedge-mean phase $\bar\chi_c$ into its local analysis basis; the physical cost is only the \emph{intra-wedge} phase dispersion,
\begin{align}
V_c^{\mathrm{pol}} = \Big|\big\langle e^{i[\chi_{\mathrm{pair}}(\varphi)
-\bar\chi_c]}\big\rangle_{\varphi\in W_c}\Big|
\;\simeq\; 1 - \frac{\big[\chi'_{\mathrm{pair}}(\varphi_c)\big]^2\,
\Delta\Phi^2}{24},
\label{eq:polvis}
\end{align}
where $\chi_{\mathrm{pair}}$ combines the signal- and idler-wedge phases of the antipodal pair and the expansion holds for smooth variation across a wedge. Polarization fidelity can be estimated using $F^{\mathrm{pol}}_c = (1+V_c^{\mathrm{pol}})/{2}$. The $\Delta\Phi^2 = (\pi/N)^2$ scaling carries a design implication that is counterintuitive: \emph{finer sectioning improves polarization uniformity}, the fidelity deficit falling as $1/N^2$, because each channel samples an ever-smaller arc of the azimuthal structure the source possesses. Segmentation is, in this respect, self-healing: the capacity limit of Eq.~\eqref{eq:capacity} is approached with the
polarization impairment vanishing quadratically. A per-wedge polarization tomography, one Bell measurement per wedge pair, on the same detector layout used for key generation, measures $\bar\chi_c$ and
$V^{\mathrm{pol}}_c$ directly and constitutes the polarization half of the calibration protocol.

\subsection{Accidentals: sectioning improves the signal-to-accidental ratio}
\label{sec:accidentals}

Accidental coincidences within channel $c$ arise from uncorrelated singles
arriving at $A_c$ and $B_c$ within the coincidence window $\tau$:
$R^{\mathrm{acc}}_c = S_{A_c} S_{B_c}\,\tau$ with per-wedge singles rates
$S_{A_c}=S_{B_c}=S_{\mathrm{tot}}/2N$. The per-channel
accidental-to-signal ratio is therefore
\begin{equation}
\frac{R^{\mathrm{acc}}_c}{R_c}
= \frac{(S_{\mathrm{tot}}/2N)^2\,\tau}{R_{\mathrm{coinc}}/N}
= \frac{S_{\mathrm{tot}}^2\,\tau}{4\,R_{\mathrm{coinc}}}\;\frac{1}{N},
\label{eq:accscaling}
\end{equation}
falling linearly with the segment count: accidentals dilute quadratically while the signal dilutes only linearly, and photons leaked into other channels' wedges never satisfy the antipodal pairing of any coincidence comparison. Sectioning therefore improves the signal-to-accidental ratio, and the improvement is convertible into rate: at a fixed per-channel accidental fraction (fixed QBER contribution), the pump power,and with it every channel's rate, may be raised in proportion to $N$. The aggregate secret key of a sectioned source is thus not merely conserved but grows with the user count until multi-pair emission within a channel, rather than accidentals, becomes the binding constraint.

\subsection{Assembling the per-channel key budget}
\label{sec:keybudget}

Collecting the sections: the per-channel QBER is
$Q_c = Q^{\mathrm{pol}}_c + Q^{\mathrm{acc}}_c$ with
$Q^{\mathrm{pol}}_c = \tfrac{2}{3}\big(1-F^{\mathrm{pol}}_c\big)$ in the
Werner reference model and $Q^{\mathrm{acc}}_c = \tfrac{1}{2}
R^{\mathrm{acc}}_c/(R_c+R^{\mathrm{acc}}_c)$; routing enters only through the rate chain of Eq.~\eqref{eq:ratechain}; and the per-channel and aggregate asymptotic secret-key rates for wavelength-independent multi-user
BBM92~\cite{BBM92, Kuniyil2026Optimizing} are
\begin{align}
R^{\mathrm{key}}_c = \big[1-(1+f)\,h(Q_c)\big]\, R_c,
\qquad \\
R^{\mathrm{key}}_{\mathrm{agg}} = \sum_{c=1}^{N} R^{\mathrm{key}}_c
\;\simeq\; \mathcal{A}(g)\,\eta\,r\,R_{\mathrm{coinc}},
\label{eq:keyrate}
\end{align}
with $h$ the binary entropy, $f$ the reconciliation inefficiency, and
$r=1-(1+f)h(Q)$ the common secret fraction when $Q_c$ is uniform,which
Eq.~\eqref{eq:polvis} guarantees to increasing accuracy as $N$ grows
\cite{DevetakWinter2005, ShorPreskill2000}. Note what
Eq.~\eqref{eq:keyrate} says about scale: the aggregate key is
$N$-independent at fixed pump, the segment count chooses how many user
pairs \emph{share} the source's key capacity, at the small guard overhead
$\mathcal{A}(g)$,while the accidentals scaling of
Eq.~\eqref{eq:accscaling} permits the pump, and hence the aggregate itself,
to grow with $N$.

A worked example closes the budget. Take the reference ring
($\theta_0=3^\circ$, $w_0=200~\mu$m, $L=3$~mm) at a collected coincidence
rate $R_{\mathrm{coinc}}=5\times10^{6}\,\mathrm{s}^{-1}$
(10~mW pump, standard collection and detection~\cite{Anwar2021}),
sectioned into $N=12$ channels with $5\%$ guards: routing consistency
$>0.9999$, acceptance $\mathcal{A}=0.88$, per-channel sifted rate
$R_c \approx 3.7\times10^{5}\,\mathrm{s}^{-1}$. With
$F^{\mathrm{pol}}=0.98$ uniform (the $1/N^2$ suppression of
Eq.~\eqref{eq:polvis} rendering azimuthal structure irrelevant at this $N$)
and accidentals at the $10^{-3}$ level, $Q_c\approx 1.4\%$ and
$r=0.77$: each of the twelve user pairs draws
$\approx 2.9\times10^{5}$ secret bits per second, an aggregate of
$\approx 3.4$~Mbit/s,$88\%$ of what the same source would deliver to a
\emph{single} user pair, now shared across twelve, with no spectral
hardware, no switching, and no per-channel source cost.
% ,,,,,,,,,, SECTION VI ,,,,,,,,,,
% ======================================================================
% SECTION VI  --  completed with REPRESENTATIVE ESTIMATES
%
% IMPORTANT: values marked with a dagger (\dag) are ASSUMED, not extracted
% from the cited papers. The pump waist of Ref. [4] is not reported, and no
% ring half-angle is stated numerically in any of the three papers; the ring
% angle of Ref. [5] is inferred from its reported 4.95 mm collimated ring at
% f = 150 mm. All derived columns are computed from these assumed inputs by
% exact pair-level Monte Carlo (the sqrt(2/pi) working form is invalid at the
% small w0q0 of Refs. [5,6]). Replace the daggered inputs with measured
% values before submission; the derived columns then follow from the same
% pipeline (azimuthal_slicing.py, guard-band branch).
% ======================================================================
%\section{Discussion and Conclusion}\label{sec:Discussion}
\section{Practical robustness}
Three practical effects warrant quantitative analysis, each incorporated into the framework via a single, measurable modification to the kernel.
\paragraph{Pump quality:} The kernel width is determined by the transverse momentum spread of the pump; a non-Gaussian or astigmatic pump replaces $\sigma_\delta = 1/w_0$ by the measured per-axis width of its angular spectrum, and an elliptic pump makes $\sigma_\varphi$ weakly azimuth-dependent—a adjustable correction, not fundamental.
\paragraph{Pointing jitter:} A pump tilt of angular scale $\sigma_{tilt}$ displaces the antipodal map by $k_p\sigma_{tilt}$ in momentum, so jitter faster than the integration time adds in quadrature, yielding $\sigma_\varphi^{\rm eff}=\sqrt{\sigma_\delta^{2}+(k_p\sigma_{\rm
tilt})^{2}}\,/\,q_0$, giving an immediate pointing specification for a target capacity; drift slower than the tracking loop is a common-mode rotation of the ring, correctable in software on an array receiver. 
\paragraph{Multi-pair emission:} The accidental advantage of Eq. \eqref{eq:polvis} invites the pump to be raised with N; the binding constraint becomes double-pair emission within a channel, which scales as the square of the per-channel rate and sets the ultimate brightness threshold of the sectioned architecture. A complete treatment belongs to the security analysis of a specific deployment; the contribution of the framework is that the threshold is per-channel and, therefore, also relaxes as $1/N$.

\section{Conclusion and extensions}
 
We have converted the azimuthal correlation of degenerate SPDC from a characterized entanglement resource into an engineering law. We have four major contributions: the exact Gaussian routing kernel $\sigma_\varphi = 1/(w_0 q_0)$, whose width no property of the crystal can
alter; the closed-form capacity $N_{\max}(\eta)\approx 3.94\,(1-\eta)\,w_0 q_0$, verified at the $10^{-3}$ level by pair-level Monte Carlo and collapsing every design question onto one dimensionless parameter; the resource theorem $N_{\max}\approx 2.2\,(1-\eta)K_{\mathrm{az}}$, which expresses the spatial channel capacity in units of azimuthal Schmidt modes; and the guard-band theorem, with its rearrangement proof of optimality, under which routing errors fall exponentially against a linear and logarithmically-growing photon cost. Around these sit the budget results,automatic rate fairness, the $1/N^2$ self-healing of azimuthal polarization structure, and the $1/N$ improvement of accidental contribution, that make the sectioned architecture more favorable as it scales. Every quantity is measurable from one far-field coincidence acquisition, and the published experiments already contain, in their fiber pitches and mirror seams, the accidental guard bands the theory sizes deliberately. Spatial-division multiplexing of entangled photons has been demonstrated repeatedly but engineered ad hoc; this framework supplies the missing design rules.

This work can be extended as follows:
\paragraph{Azimuth-Dominated Spatial Channelization.} The collinear disk ($D_0 = 0$) invites a polar grid of segments—annuli crossed with wedges—since the radial coordinate is pair-correlated through the same smearing $\sigma_\delta$. The radial capacity, however, is bounded by the disk’s momentum extent ($\propto \sqrt{k_d / L}$) divided by the routing kernel, and at millimeter crystal lengths this supports at most one radial channel at network-grade fidelity: azimuth is where the capacity lives, because only azimuth offers $2\pi$ of room at a radius that $D_0$ can increase at no fidelity cost. Short crystals or cascaded multi-ring phase matching (several discrete $D_0$ values, hence concentric rings, each an independent azimuthal circle) could add a discrete radial index, with capacities adding per ring.

\paragraph{Combining with spectral demultiplexing.} A deeper extension
exploits the nondegenerate regime. At degeneracy the spectral labels vanish and azimuth is the multiplexing resource; at nondegeneracy the radial coordinate acquires a wavelength label through the strong spatio-spectral correlation \cite{Gatti2009PRL, Kuniyil2026Entanglement} and can serve as a passive frequency demultiplexer. Because a nondegenerate ring remains azimuthally uniform and antipodally
paired at every wavelength, the two resources are compatible: a single crystal could support a two-dimensional grid of $N_{\rm az}\times N_\lambda$ user-pair channels, wedges crossed with
spectral annuli.

\paragraph{Beyond key distribution.} None of the material in Secs. \ref{sec:geometry}–\ref{sec:guard} is specific to BBM92: the capacity law and guard-band theorem govern any protocol that consumes antipodal pair channels—multiplexed heralded single-photon sources \cite{Migdall2002, Pittman2002}, parallel Bell tests and randomness generation, and entanglement-based sensor networks—wherever a ring must be shared, the same grid applies.

\section*{Acknowledgments}
The authors thank Alexander Ling for useful discussions and comments that improved this work.

\bibliography{ref} 
\end{document}